\documentclass[a4paper,oneside]{article}

\usepackage{microtype}
\usepackage[sc]{mathpazo}
\usepackage[T1]{fontenc}
\usepackage[DIV=11,BCOR=0cm,headinclude=true]{typearea}

\usepackage{amsfonts}
\usepackage{amsmath}
\usepackage{amssymb}
\usepackage{amsthm}
\usepackage{graphicx}
\usepackage{float}
\usepackage[config,labelfont=sc,textfont=sl]{caption,subfig}
\usepackage[british]{babel}
\usepackage{bm}

\usepackage[maxbibnames=100,autocite=superscript,style=custom-nature,sorting=none,defernumbers=true]{biblatex}

\usepackage{authblk}

\newcommand{\bv}[1]{\boldsymbol{#1}}
\newcommand{\bra}[1]{\langle{#1}|}
\newcommand{\ket}[1]{|{#1}\rangle}
\newcommand{\braket}[2]{\langle{#1}|{#2}\rangle}
\newcommand{\tensor}[1]{\bar{\bar{#1}}}

\DeclareMathOperator*{\imag}{Im}

\author[1,2,$\dagger$]{Parry Y.\ Chen}
\author[1]{David J.\ Bergman}
\author[2]{Yonatan Sivan}
\affil[1]{School of Physics and Astronomy, Raymond and Beverly Sackler Faculty of Exact Sciences, Tel Aviv University, Israel}
\affil[2]{Unit of Electro-optic Engineering, Ben-Gurion University, Israel}
\affil[$\dagger$]{\it{parryyu@post.bgu.ac.il}}

\title{Spectral decomposition of the Lippmann-Schwinger equation applied to cylinders}
\date{\today}

\begin{document}
\maketitle

\section{Lippmann-Schwinger equation}
When solving Maxwell's equations with sources in structured media, the most direct and common method is to solve the ``macroscopic Maxwell equations'',
\begin{align}
\nabla\times\bv{E} &= -\frac{\partial\bv{B}}{\partial t}, & \nabla\times\bv{H} &= \frac{\partial\bv{D}}{\partial t} + \bv{J}_f,
\end{align}
where $\bv{J}_f$ includes only the imposed sources and excludes bound currents due to the response of the structure. The standard wave equation is obtained,
\begin{equation}
\nabla\times(\nabla\times\bv{E}) - k^2\epsilon(\bv{r})\bv{E} = i\omega\mu_0\bv{J}_f,
\label{eq:inhomowavefree}
\end{equation}
assuming harmonic $e^{-i\omega t}$ time variation and non-magnetic media, where $k = \omega/c$. Notice that $\bv{J}_f$ is the only inhomogeneity and the effect of the structure is incorporated within the homogeneous part of the equation. One major disadvantage of the direct approach is that the Green's function,
\begin{equation}
\nabla\times(\nabla\times\tensor{G}) - k^2\epsilon(\bv{r})\tensor{G} = \tensor{I}\delta^3(\bv{r}-\bv{r}'),
\label{eq:incgreens}
\end{equation}
becomes a complicated function $\tensor{G}(\bv{r},\bv{r}')$ of nearby dielectric or plasmonic structures.

The second, and far less common, approach treats both free and bound currents equally as inhomogeneities. The advantage is that all sources remain associated with the simple Green's function for free space. This approach results in the Lippmann-Schwinger equation, which begins instead from the ``microscopic Maxwell equations'',
\begin{align}
\nabla\times\bv{E} &= -\frac{\partial\bv{B}}{\partial t}, & \nabla\times\bv{B} &= \mu_0 \left(\epsilon_0\frac{\partial\bv{E}}{\partial t} + \bv{J}\right),
\label{eq:micromax}
\end{align}
where
\begin{align}
\bv{J} &= \bv{J}_f + \bv{J}_b, & \bv{J}_b = \frac{\partial\bv{P}}{\partial t} + \nabla\times\bv{M}.
\end{align}
Restricting attention to linear non-magnetic media allows the bound current to be related to the electric field,
\begin{equation}
\bv{J}_b(\bv{r}) = -i\omega\epsilon_0(\epsilon(\bv{r})-1) \bv{E},
\end{equation}
again with harmonic time dependence. Then \eqref{eq:micromax} takes the form
\begin{equation}
\nabla\times(\nabla\times\bv{E}) - k^2\bv{E} = k^2(\epsilon(\bv{r})-1)\bv{E} + i\omega\mu_0\bv{J}_f.
\label{eq:inhomowavevac}
\end{equation}
Note that \eqref{eq:inhomowavevac} can be derived directly from \eqref{eq:inhomowavefree} via a simple manipulation. 

If the background medium is not vacuum however, it is more convenient to use a hybrid between \eqref{eq:inhomowavefree} and \eqref{eq:inhomowavevac}, taking only the difference between the structure and the background as the inhomogeneity,
\begin{equation}
\nabla\times(\nabla\times\bv{E}) - k^2\epsilon_b\bv{E} = k^2(\epsilon(\bv{r})-\epsilon_b)\bv{E} + i\omega\mu_0\bv{J}_f.
\label{eq:inhomowave}
\end{equation}
The resulting Green's function is merely scaled relative to \eqref{eq:inhomowavevac} but the RHS of \eqref{eq:inhomowave} is now non-zero only inside the structure and at external sources, which is a major advantage. This unified Green's function for all inhomogeneities is defined by
\begin{equation}
\nabla\times(\nabla\times\tensor{G}_0) - k^2\epsilon_b\tensor{G}_0 = \tensor{I}\delta^3(\bv{r}-\bv{r}'),
\end{equation}
which has a simple known analytic form $\tensor{G}_0(|\bv{r}-\bv{r}'|)$ depending on the dimensionality of the problem. 

In terms of Green's functions, the solution to \eqref{eq:inhomowave} is 
\begin{equation}
\bv{E}(\bv{r}) = \bv{E}_0(\bv{r}) + k^2 \int \tensor{G}_0 (|\bv{r} - \bv{r}'|) (\epsilon(\bv{r}') - \epsilon_b) \bv{E}(\bv{r}')\, d\bv{r}',
\label{eq:lippsch}
\end{equation}
where the integration is over source coordinates and $\bv{E}_0(\bv{r})$ is the known radiation pattern of external sources in a uniform background
\begin{equation}
\bv{E}_0(\bv{r}) = i\omega\mu_0 \int \tensor{G}_0(|\bv{r} - \bv{r}'|) \bv{J}_f(\bv{r}')\, d\bv{r}'.
\label{eq:E0}
\end{equation}
Equation \eqref{eq:lippsch} is the Lippmann-Schwinger equation for electrodynamics. Its key advantage is the simple form of the Green's function. Furthermore, a common Green's function applies to both the free sources and the response of the structure, a crucial property which we exploit to obtain an analytic solution. One apparent disadvantage is that \eqref{eq:lippsch} is an implicit equation, with the desired solution $\bv{E}(\bv{r})$ appearing inside the integral, thus forming a Fredholm integral equation of the second kind. However, by using the eigenmodes of \eqref{eq:lippsch} as a basis, this disadvantage can not only be overcome, but also grants the major advantage of enabling all possible source configurations to be simulated with the one basis set.

\section{Lippmann-Schwinger via eigenmode decomposition}
\label{sec:method}
The Lippmann-Schwinger equation, \eqref{eq:lippsch}, is the basis of two families of related numerical schemes method of moments (MoM) and discrete dipole approximation (DDA), also known as volume integral or coupled dipole methods.\autocite{purcell1973scattering,harrington1993field,lakhtakia1990macroscopic,novotny2012principles} These involves spatial discretization, recasting \eqref{eq:lippsch} in linear algebra form, which can be solved, sometimes iteratively until the solution converges.\autocite{martin1994iterative} The Lippmann-Schwinger equation can also be expanded in terms of basis functions, such as the cylindrical harmonic functions, again yielding in a linear algebra problem for the field distribution.\autocite{kristensen2010light}

We use the Lippmann-Schwinger equation as the basis of a yet more powerful analytic method.\autocite{bergman1980theory,farhi2016electromagnetic} Instead of solving \eqref{eq:lippsch} directly we solve two simpler problems: (a) the radiation pattern \eqref{eq:E0} of free sources in the absence of the structure and (b) the eigenmodes, which are the self-sustaining source-free solutions of the structure in \eqref{eq:lippsch}. The electrodynamic interaction between the source and the eigenmodes is established by exploiting the unified nature of the Green's function, applicable to both free sources and the response of the structure. The influence of the structure to the radiation pattern is expanded in terms of the eigenmodes of the structure. There are two major advantages to this method. Firstly, the simplicity of the two individual problems often admits analytic solutions. Secondly, once the eigenmodes have been obtained, the total fields are obtained immediately and analytically, for any source configuration, including position, orientation, and spatial extent, eliminating the need to repeat the simulation for each source configuration.

\subsection{The eigenvalue equation}
Consider the eigenmodes, obtained by neglecting $\bv{E}_0$ in \eqref{eq:lippsch}. At this point, we simplify the formulation by assuming that the permittivity of the structure is uniform, yielding the eigenvalue equation
\begin{equation}
s_m \bv{E}_m(\bv{r}) = k^2 \int \tensor{G}_0 (|\bv{r} - \bv{r}'|) \theta(\bv{r}') \bv{E}_m(\bv{r}')\, d\bv{r}',
\label{eq:eigen}
\end{equation}
where $s_m$ is the $m$th eigenvalue
\begin{equation}
\frac{1}{s_m} = \epsilon_m - \epsilon_b,
\label{eq:eigenvalue}
\end{equation}
and $\theta(\bv{r})$ is a function which is unity inside the structure and zero elsewhere. Note that the eigenvalue is $\epsilon_m$, representing the inclusion permittivity, which contrasts with the standard choice of eigenvalue in the literature of frequency $k$. In other words, $k$ is held fixed while $\epsilon_m$ is varied until the structure is at resonance. This leads to numerous advantages, for example, \eqref{eq:eigenvalue} demonstrates that only this choice leads to a linear eigenvalue problem, since the Green's tensor is a function of $k$ but not $\epsilon_m$. Furthermore, $k$ can be specified to be real, which yields exponentially converging rather than diverging eigenmodes.

Despite the similarity between \eqref{eq:eigen} and \eqref{eq:lippsch}, the eigenvalues $\epsilon_m$ are in general unrelated to the actual permittivity of the structure to be solved in \eqref{eq:lippsch}. Instead, the modes $\bv{E}_m$ serve as a complete orthonormal mathematical basis for expanding the fields, and the actual permittivity is specified later. The only information from \eqref{eq:lippsch} that remains in \eqref{eq:eigen} is the geometry of the inclusion, captured by $\theta(\bv{r})$. The eigenmodes are thus applicable to any uniform inclusion permittivity, even complex permittivities, without modification. While \eqref{eq:eigen} defines the eigenmodes, it is not necessary to obtain them from the integral form of the eigenvalue equation \eqref{eq:eigen}. For example, the differential form, based on \eqref{eq:inhomowave}, can be used instead, while simple structures such as infinite cylinders admit solution via the well-known step-index fiber dispersion relation, described in Section \ref{sec:disprel}.

\subsection{Eigenmode expansion}
We take as given that relatively simple task of finding the radiation pattern in the homogeneous background $\bv{E}_0$ in \eqref{eq:E0} has been completed. The final stage of the method is the rigorous solution of the Lippmann-Schwinger equation by expanding the source $\bv{J}_f(\bv{r})$ using the eigenmodes obtained from \eqref{eq:eigen}. For notational brevity, we begin by casting the Lippmann-Schwinger equation \eqref{eq:lippsch} in operator form,
\begin{equation}
\bv{E} = \bv{E}_0 + u\hat{\Gamma}\hat{\theta}\bv{E},
\label{eq:lippschgamma}
\end{equation}
where $u$ now describes the permittivity of the actual structure $\epsilon_i$, 
\begin{equation}
u = \epsilon_i-\epsilon_b.
\end{equation}
The operator $\hat{\theta}$ zeros the field outside the structure, and $\hat{\Gamma}$ is an integral operator incorporating the Green's function along with the frequency $k$,
\begin{equation}
\hat{\Gamma}\hat{\theta}\bv{E} = k^2 \int \tensor{G}_0(|\bv{r} - \bv{r}'|) \theta(\bv{r}') \bv{E}(\bv{r}')\, d\bv{r}'.
\end{equation}
Again, this represents a simplification of \eqref{eq:lippsch} to structures with a uniform permittivity, $\epsilon_i$.

The formal solution to \eqref{eq:lippschgamma} is
\begin{equation}
\bv{E} = \frac{1}{1-u\hat{\Gamma}\hat{\theta}}\bv{E}_0.
\label{eq:formal}
\end{equation}
In spectral theory, the operator $(1-u\hat{\Gamma}\hat{\theta})^{-1}$ in \eqref{eq:formal} is known as the resolvent, and the solution for the unknown field $\bv{E}$ proceeds by projecting the known $\bv{E}_0$ on to the known eigenmodes $\bv{E}_m$. We define the projection operator $\hat{I}$, which in bra-ket notation is
\begin{equation}
\hat{I} = \sum_m \hat{\theta}\ket{E_m}\bra{E_m}\hat{\theta}.
\label{eq:project}
\end{equation}
By including $\hat{\theta}$ in $\hat{I}$, we expand only over the interior fields. Firstly, this avoids an unwieldy integral over all space, and secondly the eigenmodes only provide a complete basis for expanding fields inside the structure. Note that this projection operator assumes that the modes are normalized, $\bra{E_m}\hat{\theta}\ket{E_m} = 1$. The unknown field $\ket{E}$ is then
\begin{equation}
\hat{\theta}\ket{E} = \sum_m \hat{\theta}\ket{E_m}\bra{E_m}\frac{\hat{\theta}}{1-u\hat{\Gamma}\hat{\theta}}\ket{E_0}.
\label{eq:inteqn}
\end{equation}

Next is the key step of the spectral decomposition method. Instead of applying the operator $(1-u\hat{\Gamma}\hat{\theta})^{-1}$ to $\ket{E_0}$, which would result in a lengthy numerical calculation via the Born series, we exploit the unified nature of the Green's function in \eqref{eq:lippsch} and \eqref{eq:E0} to operate on $\bra{E_m}$ instead, immediately yielding an exact analytic solution. We invoke the adjoint form of eigenvalue equation \eqref{eq:eigen},
\begin{equation}
\bra{E_m}\hat{\Gamma}\hat{\theta} = \bra{E_m}s_m.
\label{eq:eigenadj}
\end{equation}
It is critical in this step that $k$ was fixed in the definition of $s_m$ from \eqref{eq:eigenvalue}, since this ensures that the Green's tensor represented by $\hat{\Gamma}$ is identical between \eqref{eq:inteqn} and \eqref{eq:eigenadj}. This obtains from \eqref{eq:inteqn} the total interior field $\hat{\theta}\ket{E}$,
\begin{equation}
\hat{\theta}\ket{E} = \sum_m \hat{\theta}\ket{E_m} \frac{1}{1-u s_m} \bra{E_m}\hat{\theta}\ket{E_0},
\label{eq:intsol}
\end{equation}
expressed in terms of overlap integrals with the known eigenmodes. To obtain the fields everywhere, \eqref{eq:intsol} is inserted into the original Lippmann-Schwinger equation \eqref{eq:lippschgamma}, this time operating $\hat{\Gamma}\hat{\theta}$ on $\ket{E_m}$ to give
\begin{equation}
\ket{E} = \ket{E_0} + \sum_m \ket{E_m} \frac{u s_m}{1-u s_m} \bra{E_m}\hat{\theta}\ket{E_0}.
\label{eq:E0solus}
\end{equation}
For convenience, \eqref{eq:E0solus} can be rewritten explicitly in terms of permittivities,
\begin{equation}
\ket{E} = \ket{E_0} + \sum_m \ket{E_m} \frac{\epsilon_i - \epsilon_b}{\epsilon_m-\epsilon_i} \bra{E_m}\hat{\theta}\ket{E_0}.
\label{eq:E0sol}
\end{equation}

Equation \eqref{eq:E0sol} expresses the radiation of the structure in terms of its radiation in a homogeneous medium, with additional contributions from modes of the structure that are excited. The weight of each eigenmode is determined in part by the detuning between the inclusion permittivity, $\epsilon_i$, and the eigenmode, $\epsilon_m$. The eigenmode with the most similar permittivity is the dominant contributor to the radiated energy, and the series converges rapidly onto the true solution. Secondly, the electrodynamic interaction between the source and the structure is entirely encoded within the geometric factor $\bra{E_m}\hat{\theta}\ket{E_0}$, representing the spatial overlap between the incident field and the mode being excited. The explicit form of this overlap integral is presented in Section \ref{sec:adjoint}. The solution is exact up to truncation in $m$, since the Born series was avoided in obtaining \eqref{eq:intsol}, and the solution can be obtained to arbitrary accuracy by increasing $m$. The one set of eigenmodes $\ket{E_m}$ is applicable to all possible excitations $\ket{E_0}$, requiring only the evaluation of the overlap integral, which represents a small fraction of the total simulation time. Furthermore, these eigenmodes are also applicable to any inclusion permittivity, $\epsilon_i$, including lossy materials.

The form of the solution \eqref{eq:E0sol} is most suitable when the source is in the far field, so $\ket{E_0}$ has a known form, such as a plane wave or a beam. If however the source is in the near field, a second formulation is more convenient, expressed directly in terms of sources $\bv{J_f}(\bv{r})$.\autocite{farhi2016electromagnetic} This begins by casting \eqref{eq:E0} into operator form, yielding
\begin{equation}
\ket{E_0} = i\omega\mu_0 \hat{\Gamma} \ket{J_f}.
\end{equation}
After inserting into \eqref{eq:E0solus}, we obtain
\begin{equation}
\ket{E} = \ket{E_0} + i\omega\mu_0 \sum_m \ket{E_m} \frac{u s_m}{1-u s_m} \bra{E_m}\hat{\Gamma}\hat{\theta}\ket{J_f}.
\end{equation}
Again, by applying the operator $\hat{\Gamma}\hat{\theta}$ to $\bra{E_m}$ via \eqref{eq:eigenadj} rather than $\ket{J_f}$, a simple solution is obtained 
\begin{equation}
\ket{E} = \ket{E_0} + i\omega\mu_0 \sum_m \ket{E_m} \frac{u s_m^2}{1-u s_m} \braket{E_m}{J_f}.
\label{eq:Jevform}
\end{equation}
In terms of permittivities, \eqref{eq:Jevform} can be rewritten as
\begin{equation}
\ket{E} = \ket{E_0} + \frac{i}{\omega\epsilon_0} \sum_m \ket{E_m} \frac{\epsilon_i-\epsilon_b}{(\epsilon_m-\epsilon_i)(\epsilon_m-\epsilon_b)} \braket{E_m}{J_f}.
\label{eq:Jepsform}
\end{equation}
The resulting \eqref{eq:Jepsform} is largely similar to \eqref{eq:E0sol}, but the integral $\braket{E_m}{J_f}$ is now no longer restricted to the interior of the structure, and receives contributions from all locations where $\bv{J_f}(\bv{r})$ is non-zero. Nevertheless, the solution remains a rigorous solution of the Lippmann-Schwinger equation and still benefits from the completeness of the eigenmodes in expanding fields within the inclusion interior. 

\subsection{Adjoint modes}
\label{sec:adjoint}
We now give the explicit forms for the overlap integrals in \eqref{eq:E0sol} 
\begin{equation}
\bra{E_m}\hat{\theta}\ket{E_0} = \int \theta(\bv{r}) \bv{E}_m^\dagger(\bv{r}) \cdot \bv{E}_0 (\bv{r})\, d\bv{r},
\label{eq:overlapE0}
\end{equation}
and in \eqref{eq:Jepsform}, 
\begin{equation}
\braket{E_m}{J_f} = \int \bv{E}_m^\dagger(\bv{r}) \cdot \bv{J_f}(\bv{r})\, d\bv{r}.
\label{eq:overlapJ}
\end{equation}
The adjoint field $\bv{E}_m^\dagger(\bv{r})$ in \eqref{eq:overlapE0} and \eqref{eq:overlapJ} is not necessarily the complex conjugate field $\bv{E}_m^*(\bv{r})$, which is the familiar form of $\bra{E_m}$ for a self-adjoint or Hermitean operator. In fact, the adjoint field is identical to the direct field unless the structure has a symmetry. For example, an infinite cylinder has both continuous translational symmetry and continuous rotational symmetry, giving rise to $e^{i\beta z}$ and $e^{im\theta}$ variations in the respective directions. In this case, the adjoint field is obtained by the substitutions $\beta \rightarrow -\beta$ and $m \rightarrow -m$, while leaving the radial variation of the mode unchanged.\autocite{lai1990time} Alternatively, the modes may be constructed using sine and cosine linear combinations of $e^{\pm i\beta z}$ and $e^{\pm im\theta}$, which produces modes that are once again identical to their adjoint.\autocite{lai1990time}

\section{Dispersion relation and cylinder modes}
\label{sec:disprel}
To use the eigenmode decomposition formulation \eqref{eq:Jepsform}, source-free modes of the cylinder satisfying \eqref{eq:eigen} are required. These are self-sustaining modes of the structure which exist in the absense of any sources, which for cylinders can be obtained from the step-index fiber dispersion relation,
\begin{equation}
\left(\frac{1}{\alpha_m a}\frac{J'_m(\alpha_m a)}{J_m(\alpha_m a)} - \frac{1}{\alpha_b a}\frac{H'_m(\alpha_b a)}{H_m(\alpha_b a)}\right) \left(\frac{\epsilon_m}{\alpha_m a}\frac{J'_m(\alpha_m a)}{J_m(\alpha_m a)} - \frac{\epsilon_b}{\alpha_b a}\frac{H'_m(\alpha_b a)}{H_m(\alpha_b a)}\right) - \left(\frac{m\beta}{k}\right)^2 \left(\frac{1}{(\alpha_m a)^2}-\frac{1}{(\alpha_b a)^2}\right)^2 = 0,
\label{eq:disprel}
\end{equation}
where $k$ the frequency, $\beta$ is the longitudinal propagation constant, $a$ is the cylinder radius, and $\epsilon_b$ is the permittivity of the background. These variables are all pre-definied. The search variable is $\epsilon_m$, which we interpret as the permittivity of the mode, and is not the actual permittivity of the cylinder $\epsilon_i$. The in-plane propagation constants are given by
\begin{align}
\alpha_m^2 &= k^2\epsilon_m - \beta^2, & \alpha_b^2 &= k^2\epsilon_b - \beta^2.
\label{eq:alpha}
\end{align}

\begin{figure}[!t]
\begin{center}
\subfloat{\includegraphics{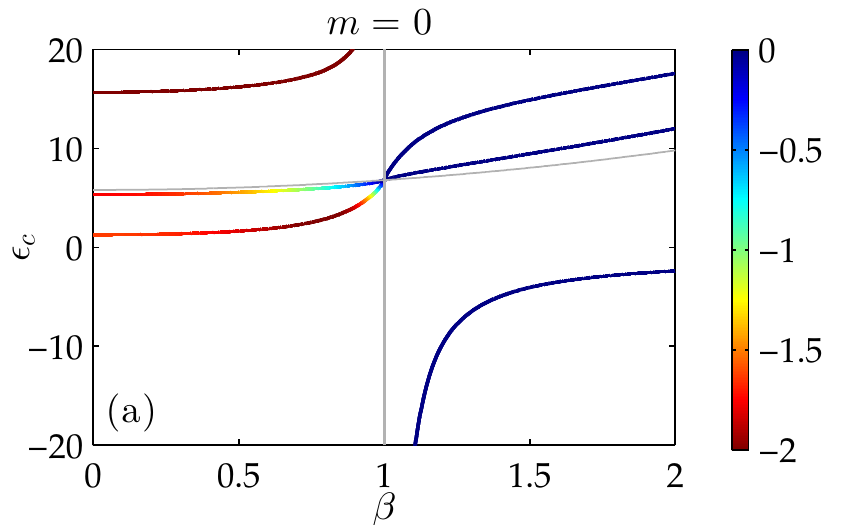}}\\
\noindent\makebox[\textwidth]{%
\subfloat{\includegraphics{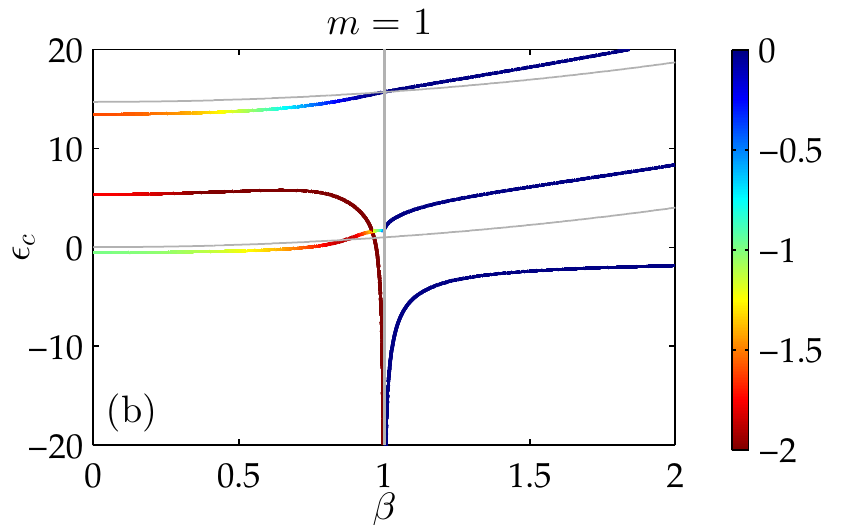}}
\subfloat{\includegraphics{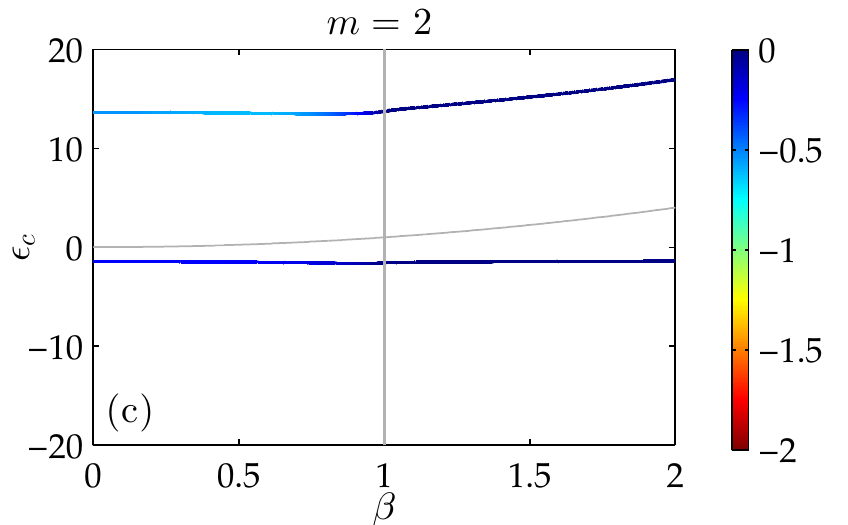}}
}
\caption[]{Dispersion relations with complex mode permittivity $\epsilon_m$ against propagation constant $\beta$, with $\imag(\epsilon_m)$ indicated by color. Wavenumber is $k=1$ and cylinder radius $a$ is normalized to 1. The subfigures plot modes with a common angular order, given by the title, of differing radial orders. The vertical gray line represents the light line, while the curved gray lines are singularities of the dispersion relation \eqref{eq:disprel}. Bound modes with real $\epsilon_m$ exist to the right of the light line, while complex $\epsilon_m$ radiative modes exist to the left.}
\label{fig:disp}
\end{center}
\end{figure}

The cylinder dispersion relation \eqref{eq:disprel} is a transcendental equation, for which an efficient and robust solution is available via contour integration methods.\autocite{chen2017robust} Examples of dispersion relations obtained from \eqref{eq:disprel} with $\epsilon_m$ as the eigenvalue are plotted in Figure \ref{fig:disp}. Such modes differ from the modes typically sought from the fiber dispersion relation \eqref{eq:disprel}, where $\beta$ is the eigenvalue. Here, we typically specify real $k$ and $\beta$, and assuming a lossless background $\epsilon_b$, two distinct regimes exist on the two sides of the light line. When $k^2\epsilon_b < \beta^2$, $\alpha_b^2$ is real and negative, and the modes decay exponentially towards infinity, thereby satisfying Dirichlet boundary conditions. When $k^2\epsilon_b > \beta^2$, $\alpha_b^2$ is real and positive, corresponding to outgoing fields that radiate energy towards infinity, thus satisfying Sommerfeld boundary conditions. These modes do not exponentially diverge at infinity, unlike the corresponding modes above the light line when $\beta$ is the eigenvalue. To maintain the steady state despite the outflux of energy, $\epsilon_m$ is complex and represents a cylinder with gain. 

Due to the symmetry of the infinite cylinder, the modes can be identified by the propagation constant $\beta$ and two additional quantum numbers: $m$ the azimuthal order, and $l$ the radial order, which count the number of nodes in the respective directions. Different orders $m$ are solutions to different dispersion relations \eqref{eq:disprel}, while different orders $l$ are different solutions to the same dispersion relation. Orders $m$ extend from $-\infty$ to $\infty$, though only orders $m \geqslant 0$ need to be found due to symmetry. Radial orders $l$ are numbered from $0$ to $\infty$. For the case of a positive index background, usually a single plasmonic mode exists with with negative $\epsilon_m$, denoted by $l=0$, while an infinite number of dielectric modes with positive $\epsilon_m$ exist.

The specific field profiles are determined by the solutions of \eqref{eq:disprel}. In the background, outgoing fields exist in the absence of any incoming fields
\begin{align}
E_{m,z} &= B_m^E H_m(\alpha_b r) e^{im\theta} e^{i\beta z}, &
H_{m,z} &= B_m^H H_m(\alpha_b r) e^{im\theta} e^{i\beta z},
\label{eq:extfields}
\end{align}
while fields inside the cylinder have the form
\begin{align}
E_{m,z} &= C_m^E J_m(\alpha_m r) e^{im\theta} e^{i\beta z}, &
H_{m,z} &= C_m^H J_m(\alpha_m r) e^{im\theta} e^{i\beta z}.
\end{align}
The ratios between the coefficients $B_m^E$, $B_m^H$, $C_m^E$, and $C_m^H$ are fixed. The in-plane field components can be derived using the expressions in Appendix \ref{sec:cylcoords}.

\appendix
\section{Cylindrical harmonic fields}
The most natural basis for all cylindrical geometries is the cylindrical harmonic basis. However, there are two major variants. Firstly, fields may be specified by their longitudinal components $E_z$ and $H_z$, from which all other fields components $E_r$, $E_\theta$, $H_r$, and $H_\theta$ can be obtained. Appendix \ref{sec:cylcoords} treats this representation, in particular obtaining the four in-plane components from the two longitudinal components. The second representation uses the rotational basis, which expresses the in-plane fields components of the electric field using the eigenvectors of the rotation operator. This representation is ideal for the analytic evaluation of the overlap and normalization integrals of Appendix \ref{sec:integrals}, which involve dot products between the three components of the electric field. Appendix \ref{sec:rotbasis} treats this representation.

\subsection{Fields in cylindrical coordinates}
\label{sec:cylcoords}
Fields in the cylindrical harmonic basis specified by their longitudinal components have the form
\begin{align}
E_z &= \sum_m C_m^E \mathcal{Z}_m(\alpha r) e^{im\theta} e^{i\beta z}, &
H_z &= \sum_m C_m^H \mathcal{Z}_m(\alpha r) e^{im\theta} e^{i\beta z},
\label{eq:Ez}
\end{align}
where $\mathcal{Z}_m(z)$ is any linear combination of Bessel functions. The infinite translational symmetry along $z$ allows the perpendicular components to be found,
\begin{align}
\bv{E}_\perp &= \frac{i}{\alpha^2}(\beta \nabla_\perp E_z - k\mu\bv{\hat{z}}\times\nabla_\perp H_z), &
\bv{H}_\perp &= \frac{i}{\alpha^2}(k\epsilon\bv{\hat{z}}\times\nabla_\perp E_z + \beta\nabla_\perp H_z).
\label{eq:inplane}
\end{align}
The magnetic field $\bv{H}$ has been rescaled by the free space impedance, $\bv{H} = \sqrt{\mu_0/\epsilon_0}\tilde{\bv{H}}$, where $\tilde{\bv{H}}$ is the SI quantity. Harmonic $e^{-i\omega t}$ time dependence was assumed. This expression can be derived from Maxwell's equations.

The explicit polar form of \eqref{eq:inplane} can be obtained from the gradient operator,
\begin{equation}
\nabla_\perp = \bv{\hat{r}}\frac{\partial}{\partial r} + \bv{\hat{\theta}}\frac{1}{r}\frac{\partial}{\partial\theta},
\end{equation}
to yield
\begin{align}
E_r &= \frac{i}{\alpha^2}\left(\beta\frac{\partial}{\partial r}E_z + k\mu\frac{1}{r}\frac{\partial}{\partial\theta}H_z\right), &
H_r &= \frac{i}{\alpha^2}\left(\beta\frac{\partial}{\partial r}H_z - k\epsilon\frac{1}{r}\frac{\partial}{\partial\theta}E_z\right), \label{eq:cylr}\\
E_\theta &= \frac{i}{\alpha^2}\left(\frac{\beta}{r}\frac{\partial}{\partial\theta}E_z - k\mu\frac{\partial}{\partial r}H_z\right), &
H_\theta &= \frac{i}{\alpha^2}\left(\frac{\beta}{r}\frac{\partial}{\partial\theta}H_z + k\epsilon\frac{\partial}{\partial r}E_z\right). \label{eq:cylth}
\end{align}

Applying the cylindrical form \eqref{eq:cylr}--\eqref{eq:cylth} derives the in-plane fields from \eqref{eq:Ez},
\begin{align}
E_r &= \sum_m \left[C_m^E \frac{i\beta}{\alpha} \mathcal{Z}_m'(\alpha r) - C_m^H \frac{mk\mu}{r\alpha^2} \mathcal{Z}_m(\alpha r)\right] e^{im\theta} e^{i\beta z}, \label{eq:Er}\\
E_\theta &= \sum_m \left[-C_m^E \frac{m\beta}{r\alpha^2} \mathcal{Z}_m(\alpha r) - C_m^H \frac{ik\mu}{\alpha} \mathcal{Z}_m'(\alpha r)\right] e^{im\theta} e^{i\beta z}, \label{eq:Et}\\
H_r &= \sum_m \left[C_m^H \frac{i\beta}{\alpha} \mathcal{Z}_m'(\alpha r) + C_m^E\frac{mk\epsilon}{r\alpha^2} \mathcal{Z}_m(\alpha r)\right] e^{im\theta} e^{i\beta z},\\
H_\theta &= \sum_m \left[-C_m^H \frac{m\beta}{r\alpha^2} \mathcal{Z}_m(\alpha r) + C_m^E \frac{ik\epsilon}{\alpha} \mathcal{Z}_m'(\alpha r)\right] e^{im\theta} e^{i\beta z},
\end{align}
where $\mathcal{Z}_m'(z)$ denotes differentiation with respect to the argument. 

\subsection{Fields in rotational basis}
\label{sec:rotbasis}
The in-plane field components can also be expressed in the rotational basis, useful for calculating overlap integrals, given by 
\begin{align}
\bv{\hat{e}}_+ &= \frac{1}{\sqrt{2}}(\bv{\hat{x}} + i\bv{\hat{y}}), & \bv{\hat{e}}_- &= \frac{1}{\sqrt{2}}(\bv{\hat{x}} - i\bv{\hat{y}}).
\label{eq:rotbasis}
\end{align}
These are the eigenvectors of the rotation operator
\begin{equation}
\begin{bmatrix} \cos\theta & \sin\theta \\ -\sin\theta & \cos\theta \end{bmatrix}
\begin{bmatrix} \bv{\hat{x}} \\ \bv{\hat{y}} \end{bmatrix} = 
\lambda \begin{bmatrix} \bv{\hat{x}} \\ \bv{\hat{y}} \end{bmatrix}.
\end{equation}
The identities \eqref{eq:rotbasis} can also be written in terms of $\bv{\hat{r}}$ and $\bv{\hat{\theta}}$, 
\begin{align}
\bv{\hat{e}}_+ &= \frac{1}{\sqrt{2}}(e^{i\theta}\bv{\hat{r}} + ie^{i\theta}\bv{\hat{\theta}}), & 
\bv{\hat{e}}_- &= \frac{1}{\sqrt{2}}(e^{-i\theta}\bv{\hat{r}} - ie^{-i\theta}\bv{\hat{\theta}}).
\label{eq:rotbasiscyl}
\end{align}

Using \eqref{eq:rotbasiscyl} to convert \eqref{eq:Ez}, \eqref{eq:Er}, and \eqref{eq:Et}, the total electric field has a particularly simple form,
\begin{equation}
\begin{split}
\bv{E} &= \sum_m\left[\bv{\hat{z}} C_m^E \mathcal{Z}_m(\alpha r)e^{im\theta} - \bv{\hat{e}}_+ \frac{1}{\sqrt{2}\alpha}(i\beta C_m^E + k\mu C_m^H)\mathcal{Z}_{m+1}(\alpha r)e^{i(m+1)\theta}\right.\\
&+ \left.\bv{\hat{e}}_-\frac{1}{\sqrt{2}\alpha}(i\beta C_m^E - k\mu C_m^H)\mathcal{Z}_{m-1}(\alpha r)e^{i(m-1)\theta}\right] e^{i\beta z},
\end{split}
\label{eq:Erotbasis}
\end{equation}
which was simplified using the Bessel function identities
\begin{align}
\frac{m}{z}\mathcal{Z}_m(z) &= \frac{1}{2}(\mathcal{Z}_{m-1}(z) + \mathcal{Z}_{m+1}(z)), &
\mathcal{Z}_m'(z) = \frac{1}{2}(\mathcal{Z}_{m-1}(z) - \mathcal{Z}_{m+1}(z)).
\label{eq:bessid}
\end{align}
The magnetic field can also be cast in the rotational basis, but we shall have no need for such an expression, since all necessary overlap integrals are expressed entirely in terms of electric fields.

\section{Overlap and normalization integrals}
\label{sec:integrals}
The practical use of eigenmode decomposition method involves two key steps: finding the eigenmodes, and then using the eigenmodes as a basis for projection. The latter requires evaluation of several integrals, such as normalization and the overlap integrals \eqref{eq:overlapE0}. These can be performed analytically for cylindrical geometries, and have the common form yielding a general result derived in Appendix \ref{sec:genint} and \ref{sec:bessints}, applied to normalization integrals in \ref{sec:norm}.

\subsection{General result}
\label{sec:genint}
Consider the integral
\begin{equation}
\mathcal{I} = \int \bv{E}_1^\dagger \cdot \bv{E}_2\, dA
\label{eq:intdef}
\end{equation}
evaluated over the circular domain of the inclusion, where the dagger represents the adjoint of the mode. This typically represents the projection of a general field $\bv{E}_2$ onto a mode $\bv{E}_1$, though in the case where $\bv{E}_1$ and $\bv{E}_2$ are equal this represents a normalization integral. The adjoint mode satisfies the same governing differential equation as the direct mode, but with the opposite propagation constants in any coordinate where the structure exhibits symmetry. This entails the substitution $m \rightarrow -m$ and $\beta \rightarrow -\beta$ in \eqref{eq:Ez}--\eqref{eq:Et}. 

Since $\bv{E}_1$ usually represents a single eigenmode in the interior of the cylinder, it is represented by Bessel functions of the first kind with only a single pre-defined azimuthal order $m$. To obtain the adjoint field, we follow the same procedure which yielded \eqref{eq:Erotbasis}, to obtain
\begin{equation}
\bv{E}_1^\dagger = [\bv{\hat{z}} C_{1,m}^E J_m(\alpha_1 r)e^{-im\theta} + \bv{\hat{e}}_+ C_{1,m}^{\dagger,+} J_{m-1}(\alpha_1 r)e^{-i(m-1)\theta} + \bv{\hat{e}}_- C_{1,m}^{\dagger,-} J_{m+1}(\alpha_1 r)e^{-i(m+1)\theta}] e^{-i\beta z},
\label{eq:E1adj}
\end{equation}
with coefficients
\begin{align}
C_{1,m}^{\dagger,+} &= \frac{1}{\sqrt{2} \alpha_1}(-i\beta C_{1,m}^E + k\mu C_{1,m}^H), &
C_{1,m}^{\dagger,-} &= \frac{1}{\sqrt{2} \alpha_1}(i\beta C_{1,m}^E + k\mu C_{1,m}^H).
\end{align}
The successful analytic evaluation of \eqref{eq:intdef} requires that the cylindrical harmonic basis for $\bv{E}_2$ share the same origin as $\bv{E}_1^\dagger$. This can be achieved with Graf's addition theorem for example, ensuring the fields are also expressed as Bessel functions of the first kind but with all azimuthal orders $n$,
\begin{equation}
\bv{E}_2 = \sum_{n=-\infty}^{\infty} [\bv{\hat{z}} C_{2,n}^E J_n(\alpha_2 r)e^{in\theta} + \bv{\hat{e}}_+ C_{2,n}^+ J_{n+1}(\alpha_2 r)e^{i(n+1)\theta} + \bv{\hat{e}}_- C_{2,n}^- J_{n-1}(\alpha_2 r)e^{i(n-1)\theta}] e^{i\beta z},
\label{eq:E2}
\end{equation}
with coefficients as in \eqref{eq:Erotbasis},
\begin{align}
C_{2,m}^+ &= -\frac{1}{\sqrt{2} \alpha_2}(i\beta C_{2,m}^E + k\mu C_{2,m}^H), &
C_{2,m}^- &= \frac{1}{\sqrt{2} \alpha_2}(i\beta C_{2,m}^E - k\mu C_{2,m}^H).
\end{align}

We now evaluate the integral \eqref{eq:intdef} in general form with fields \eqref{eq:E1adj} and \eqref{eq:E2}. With the appropriate coefficients $C_{2,m}^E$ and $C_{2,m}^H$, this general form can be particularized to various kinds of overlap integrals and the normalization integral. To begin, first note the unusual orthogonality relation obeyed by the rotational basis vectors,
\begin{align}
\bv{\hat{e}}_+ \cdot \bv{\hat{e}}_+ &= 0, & \bv{\hat{e}}_- \cdot \bv{\hat{e}}_- &= 0, & \bv{\hat{e}}_+ \cdot \bv{\hat{e}}_- &= 1.
\end{align}
Thus, a general dot product in this basis has 3 contributions when expanded,
\begin{equation}
(A_z\bv{\hat{z}} + A_+\bv{\hat{e}}_+ + A_-\bv{\hat{e}}_-) \cdot (B_z\bv{\hat{z}} + B_+\bv{\hat{e}}_+ + B_-\bv{\hat{e}}_-) = A_z B_z + A_+ B_- + A_- B_+.
\end{equation}

The first contribution to \eqref{eq:intdef} is
\begin{equation}
\begin{split}
\mathcal{I}_{\bv{\hat{z}}\cdot\bv{\hat{z}}} &= \int_0^{2\pi} \int_0^a C_{1,m}^E J_m(\alpha_1 r) e^{-im\theta} \sum_n C_{2,n}^E J_n(\alpha_2 r) e^{in\theta}\, r\, dr\, d\theta\\
&= 2\pi C_{1,m}^E C_{2,m}^E \int_0^a J_m(\alpha_1 r) J_m(\alpha_2 r)\, r\, dr\, d\theta\\
&= 2\pi C_{1,m}^E C_{2,m}^E I_m(\alpha_1, \alpha_2, a),
\end{split}
\end{equation}
where $a$ is the radius of the cylinder and defining the integral that appears as
\begin{equation}
I_m(\alpha_1, \alpha_2, a) = \int_0^a J_m(\alpha_1 r) J_m(\alpha_2 r)\, r\, dr.
\label{eq:bessint}
\end{equation}
These integrals can be evaluated analytically, using procedures described in Appendix \ref{sec:bessints}. The other two contributions to \eqref{eq:intdef} proceed in a similar fashion,
\begin{equation}
\begin{split}
\mathcal{I}_{\bv{\hat{e}}_+\cdot\bv{\hat{e}}_-} &= \int_0^{2\pi} \int_0^a C_{1,m}^{\dagger,+} J_{m-1}(\alpha_1 r) e^{-i(m-1)\theta} \sum_n C_{2,n}^- J_{n-1}(\alpha_2 r) e^{i(n-1)\theta}\, r\, dr\, d\theta\\
&= 2\pi C_{1,m}^{\dagger,+} C_{2,m}^- I_{m-1}(\alpha_1, \alpha_2, a),
\end{split}
\end{equation}
and
\begin{equation}
\begin{split}
\mathcal{I}_{\bv{\hat{e}}_-\cdot\bv{\hat{e}}_+} &= \int_0^{2\pi} \int_0^a C_{1,m}^{\dagger,-} J_{m+1}(\alpha_1 r) e^{-i(m+1)\theta} \sum_n C_{2,n}^+ J_{n+1}(\alpha_2 r) e^{i(n+1)\theta}\, r\, dr\, d\theta\\
&= 2\pi C_{1,m}^{\dagger,-} C_{2,m}^+ I_{m+1}(\alpha_1, \alpha_2, a),
\end{split}
\end{equation}
These three contributions give in total
\begin{equation}
\mathcal{I} = 2\pi[C_{1,m}^E C_{2,m}^E I_m(\alpha_1, \alpha_2, a) + C_{1,m}^{\dagger,+} C_{2,m}^- I_{m-1}(\alpha_1, \alpha_2, a) + C_{1,m}^{\dagger,-} C_{2,m}^+ I_{m+1}(\alpha_1, \alpha_2, a)].
\label{eq:cylintsol}
\end{equation}

\subsection{Bessel integrals}
\label{sec:bessints}
Explicit expressions for the integral \eqref{eq:bessint} are now obtained. This integral appears in Abramowitz and Stegun (11.3.29),
\begin{equation}
I_m(\alpha_1, \alpha_2, a) = \frac{\alpha_1 a J_{m+1}(\alpha_1 a) J_m(\alpha_2 a)  - \alpha_2 a J_{m+1}(\alpha_2 a) J_m(\alpha_1 a)}{\alpha_1^2 - \alpha_2^2}.
\label{eq:bessintsol}
\end{equation}
However, this expression cannot be used when propagation constants match, $\alpha_1 = \alpha_2$, which occurs when normalization integrals are being evaluated. Instead, the following form applies,
\begin{equation}
I(\alpha,\alpha,a) = \frac{a^2}{2} [J_m(\alpha a)^2 - J_{m+1}(\alpha a) J_{m-1}(\alpha a)].
\label{eq:bessintalt}
\end{equation}
The expression \eqref{eq:bessintalt} can be derived from \eqref{eq:bessintsol} by taking the limit $\alpha_2 \rightarrow \alpha_1$, i.e.\ by considering $I(\alpha,\alpha,a) \approx I(\alpha, \alpha+\varepsilon, a)$ and retaining first order terms. This proceeds from the alternative form of \eqref{eq:bessintsol},
\begin{equation}
I_m(\alpha_1, \alpha_2, a) = \frac{\alpha_2 a J'_m(\alpha_2 a) J_m(\alpha_1 a) - \alpha_1 a J'_m(\alpha_1 a) J_m(\alpha_2 a) }{\alpha_1^2 - \alpha_2^2},
\end{equation}
obtained using the recurrence relation for Bessel functions,
\begin{equation}
\mathcal{Z}'_m(z) = -\mathcal{Z}_{m+1}(z) + \frac{m}{z}\mathcal{Z}_m(z).
\end{equation}

The derivation begins with
\begin{equation}
I(\alpha,\alpha,a) = \frac{(\alpha+\varepsilon) a J'_m((\alpha+\varepsilon)a) J_m(\alpha a) - \alpha a J'_m(\alpha a) J_m((\alpha+\varepsilon)a)}{\alpha^2 - (\alpha+\varepsilon)^2}.
\end{equation}
The Bessel functions are expanded in terms of its derivatives,
\begin{align}
J_m(\alpha a + \varepsilon a) &\approx J_m(\alpha a) + J'_m(\alpha a)\varepsilon a + \frac{1}{2} J''_m(\alpha a)(\varepsilon a)^2 + \cdots\\
J'_m(\alpha a + \varepsilon a) &\approx J'_m(\alpha a) + J''_m(\alpha a)\varepsilon a + \frac{1}{2} J'''_m(\alpha a)(\varepsilon a)^2 + \cdots,
\end{align}
and retaining only leading order terms in $\varepsilon$, we obtain
\begin{equation}
I(\alpha,\alpha,a) \approx \frac{(\alpha a)^2 J_m(\alpha a) J''_m(\alpha a)\varepsilon + \alpha a J_m(\alpha a) J'_m(\alpha a)\varepsilon - (\alpha a)^2 J'_m(\alpha a)J'_m(\alpha a)\varepsilon}{2\alpha^2\varepsilon}.
\end{equation}
This may be simplified using the defining Bessel differential equation
\begin{equation}
z^2\mathcal{Z}''_m(z) + z\mathcal{Z}'(z) + (z^2 - m^2)\mathcal{Z}_m(z) = 0.
\end{equation}
Thus,
\begin{align}
I(\alpha,\alpha,a) &= \frac{1}{2\alpha^2} [((\alpha a)^2 - m^2) J_m(\alpha a)^2 + (\alpha a)^2 J'_m(\alpha a)^2]\\
&= \frac{a^2}{2} \left[J_m(\alpha a)^2 + J'_m(\alpha a)^2 - \left(\frac{m}{\alpha a}\right)^2 J_m(\alpha a)^2\right]\\
&= \frac{a^2}{2} [J_m(\alpha a)^2 - J_{m+1}(\alpha a) J_{m-1}(\alpha a)],
\end{align}
which was simplified using the Bessel identities \eqref{eq:bessid}.

\subsection{Normalization integral}
\label{sec:norm}
The simplest application of \eqref{eq:cylintsol} occurs during evaluation of the normalization integral, which is necessary for successful projection \eqref{eq:project},
\begin{equation}
\mathcal{N}^2 = \int \bv{E}_m^\dagger \cdot \bv{E}_m\, dA,
\label{eq:normint}
\end{equation}
where fields are expressed as
\begin{align}
\bv{E}_m &= \bv{\hat{z}} C_m^E J_m(\alpha_m r)e^{im\theta} + \bv{\hat{e}}_+ C_m^+ J_{m-1}(\alpha_m r)e^{i(m-1)\theta} + \bv{\hat{e}}_- C_m^- J_{m+1}(\alpha_m r)e^{i(m+1)\theta}, \\
\bv{E}_m^\dagger &= \bv{\hat{z}} C_m^E J_m(\alpha_m r)e^{-im\theta} + \bv{\hat{e}}_+ C_m^{\dagger,+} J_{m-1}(\alpha_m r)e^{-i(m-1)\theta} + \bv{\hat{e}}_- C_m^{\dagger,-} J_{m+1}(\alpha_m r)e^{-i(m+1)\theta}, \label{eq:modefieldadj}
\end{align}
suppressing the longitudinal variation. The coefficients are
\begin{align}
C_m^+ &= -\frac{1}{\sqrt{2} \alpha_m}(i\beta C_m^E + k\mu C_m^H), &
C_m^- &= \frac{1}{\sqrt{2} \alpha_m}(i\beta C_m^E - k\mu C_m^H), \\
C_m^{\dagger,+} &= \frac{1}{\sqrt{2} \alpha_m}(-i\beta C_m^E + k\mu C_m^H), &
C_m^{\dagger,-} &= \frac{1}{\sqrt{2} \alpha_m}(i\beta C_m^E + k\mu C_m^H).
\label{eq:modecoeffadj}
\end{align}
The result analytic result for \eqref{eq:normint} is
\begin{equation}
\mathcal{N}^2 = 2\pi[(C_m^E)^2 I_m(\alpha_m,\alpha_m,a) + C_m^+ C_m^{\dagger,-} I_{m+1}(\alpha_m,\alpha_m,a) + C_m^- C_m^{\dagger,+} I_{m-1}(\alpha_m,\alpha_m,a)],
\end{equation}
which can be evaluated using \eqref{eq:bessintalt}.

\printbibliography
\end{document}